\documentclass[12pt]{article}

\usepackage{amsfonts}

\usepackage{amssymb}
\usepackage{latexsym}
\usepackage{graphicx}
\usepackage[english]{babel}

\usepackage{amsfonts}
\usepackage{latexsym}
\usepackage{graphicx}
\usepackage[english]{babel}
\begin{document}
\input epsf

\def\p{\partial}
\def\h{{1\over 2}}
\def\be{\begin{equation}}
\def\bea{\begin{eqnarray}}
\def\ee{\end{equation}}
\def\eea{\end{eqnarray}}
\def\d{\partial}
\def\la{\lambda}
\def\eps{\epsilon}
\def\bb{\bigskip}
\def\mm{\medskip}
\newcommand{\dm}{\begin{displaymath}}
\newcommand{\edm}{\end{displaymath}}
\renewcommand{\b}{\tilde{B}}
\newcommand{\gm}{\Gamma}
\newcommand{\ac}[2]{\ensuremath{\{ #1, #2 \}}}
\renewcommand{\ell}{l}
\newcommand{\z}{\ell}
\newcommand{\newsection}[1]{\section{#1} \setcounter{equation}{0}}
\def\bb{$\bullet$}
\def\Qbar{{\bar Q}_1}
\def\QPbar{{\bar Q}_p}

\def\q{\quad}

\def\bn{B_\circ}

\let\a=\alpha \let\b=\beta \let\g=\gamma \let\d=\delta \let\e=\epsilon
\let\c=\chi \let\th=\theta  \let\k=\kappa
\let\l=\lambda \let\m=\mu \let\n=\nu \let\x=\xi \let\r=\rho
\let\s=\sigma \let\t=\tau
\let\vp=\varphi \let\vep=\varepsilon
\let\w=\omega      \let\G=\Gamma \let\D=\Delta \let\Th=\Theta
                     \let\P=\Pi \let\S=\Sigma

\def\h{{1\over 2}}
\def\t{\tilde}
\def\r{\rightarrow}
\def\nn{\nonumber\\}
\let\bm=\bibitem
\def\Kt{{\tilde K}}
\def\b{\bigskip}

\let\p=\partial

\begin{flushright}
%OHSTPY-HEP-T-03-012\\
\end{flushright}
\vspace{20mm}
\begin{center}
{\LARGE  What can the information paradox tell us about the early Universe?\footnote{Essay awarded second prize in the Gravity Research Foundation essay competition 2012}}
\\
\vspace{18mm}
{\bf  Samir D. Mathur }\\

\vspace{8mm}
Department of Physics,\\ The Ohio State University,\\ Columbus,
OH 43210, USA\\mathur@mps.ohio-state.edu
\vspace{4mm}
\end{center}
\vspace{10mm}
\thispagestyle{empty}
\begin{abstract}

In recent years we have come to understand how the information paradox is resolved in string theory. The huge entropy $S_{bek}={A\over 4G}$ of black holes is realized by an explicit set of horizon sized `fuzzball' wavefunctions. The wavefunction of a collapsing shell spreads relatively quickly over this large phase space of states, invalidating the classical black hole geometry the shell would have created. We argue that a related effect may occur in the early Universe. When matter is crushed to high densities we can access a similarly large phase space of gravitational `fuzzball' solutions. While we cannot estimate specific  quantities at this point, a qualitative analysis suggests that spreading over phase space creates  an extra `push' expanding the Universe to larger volumes. 

\end{abstract}
\vskip 1.0 true in

\newpage
\setcounter{page}{1}

Consider a toroidal box of volume $V$. In this box, put  energy $E$. What is the nature of the state  in the limit $E\r \infty$? In particular, what is the entropy $S(E,V)$?

This question has never been seriously addressed in string theory which aims to be a complete theory including gravity. The reason can be traced to a general feeling: Won't the maximal entropy state be a {\it black hole}, whose states we cannot really understand? 

But this is where the question gets interesting. Suppose we take
\be
V=(1~{\rm meter})^3, ~~~E=1~M_\odot
\ee
A solar mass black hole has radius $3~{\rm Km}$, so it cannot fit into our meter sized box.  There are four possibilities:

\b

(a) The entropy in the box will be that of a solar mass hole

(b) The entropy will be that of a hole of size 1 meter, the largest that can fit in our box

(c) The entropy will be something else, perhaps in between (a) and (b)

(d) We cannot put one solar mass of energy in the box of size 1 meter

\b

To see why this question is relevant to Cosmology, forget for a moment the notions of inflation and dark energy,  and imagine following a 
traditional big bang Cosmology to early times $t\r 0$. Let the $t=const$  slice be a torus; thus we have a flat Universe with torus volume $V$ shrinking towards zero as $t\r 0$.  What will the torus contain?

At large $t$, the torus contains dust. At smaller $t$, we expect radiation, which has an entropy $S\sim E^{3\over 4}$. What happens at even smaller $t$ depends on the theory; for example in string theory we will enter a `string gas' phase where $S\sim E$. Proceeding further back, we ask: What is the nature of the maximal entropic state in the limit of infinite energy density?

\b

\b

\begin{figure}[htbp]
\begin{center}
\includegraphics[scale=.38]{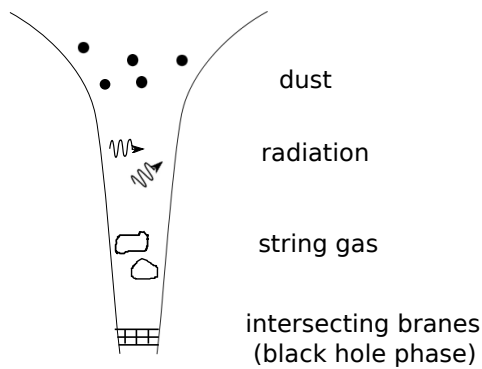}
\caption{{Different states of  string theory as we follow the Universe back to early times.}}
\label{ftwo}
\end{center}
\end{figure}

\b

\b

Two clarifications before we proceed:

\b

(i) One may wonder if we should look for {\it low} entropy states so that the entropy can increase in the future. But in fact we have traditionally considered the {\it maximal} entropy state at any given $t$: in the traditional big bang we took the radiation phase at early times because radiation has more entropy than dust, and we take a string gas phase at even earlier times because the string gas has more entropy than radiation. The point is that we can have the maximal entropy state at a {\it given}  volume $V$; when $V$ increases the entropy can increase further, with a change in the nature of the state from string gas to radiation to dust.  

\b

(ii) Should we consider the maximal entropy state if it turns out to involve black holes \cite{bek}? If the Universe started off with black holes then would it not always have these holes, in contradiction to what we see around us today? But in recent years we have understood that black hole states are no different from other states in string theory. If we switch off gravity, then the entropy of black holes is reproduced by the states of multiply intersecting branes \cite{sv}. If we turn on gravity, these states  `swell up' to become `fuzzballs', with the following geometric structure \cite{fuzzball}. The 6 small  compact directions of string theory make nontrivial fibrations over the 3 space directions in the volume $V$,  creating generalized analogues of Kaluza-Klein monopoles and antimonopoles. The set of all such solutions account for the Bekenstein entropy of the hole \cite{rychkov}.  We will not need the detailed structure of these solutions; all we note is that black hole states are not fundamentally different from other states in string theory.

\b

\b

 In standard general relativity there is no constraint on how much $E$ we can put in a given volume $V$; more $E$ simply gives a larger rate of expansion, via the relation $\Big ({\dot a\over a}\Big )^2={8\pi G\over 3} \rho$. Thus we ignore (d). To choose between the other possibilities we must look into the structure of  microstates. A study of simple fuzzballs  suggests that we get (b). To see the reason we recall why the typical fuzzball has size of order the Schwarzschild radius. When we explicitly construct the states of the hole, we  get $e^{S_{bek}}$ orthogonal wavefunctions of the string fields. In order to be orthogonal the different wavefunctions cannot really overlap, and the minimum region requires to hold all the wavefunctions turns out to be order the Schwarzschild radius.

Now if we limit the size of our region to be 1 meter (while still taking one solar mass of energy), then not all the $e^{S_{bek}}$ wavefunctions for the solar mass hole can fit in our box. An explicit study of the wavefunctions of the 2-charge extremal hole indicates that possibility (b) holds in that case \cite{phase}, and in what follows we will assume that (b) is correct for more general situations as well.

The above conclusion may seem natural in retrospect: the entropy $S(E)$ in a volume is order the entropy of the largest hole which can fit in that volume. But note that we have gone somewhat beyond the standard Bekenstein relation $S={A\over 4G}$ \cite{bek}: there is no boundary in the torus that we can compute the area of, and the energy $E$ is {\it more} than that of the hole that would fit in the given volume $V$.

To see how the above observation may be important for Cosmology, let us  recall how the information paradox \cite{hawking} gets resolved in string theory. Consider a shell of mass $M$ that is collapsing through its horizon radius $R\sim GM$. In ordinary 3+1 dimensional gravity  the wavefunction of the shell moves in the expected way to smaller $r$, creating the traditional black hole geometry. But in string theory the $e^{S_{bek}}$ states of the hole give alternate wavefunctions with the same quantum numbers as the shell. There is a small amplitude for the wavefunction of the shell to tunnel into one of these microstate wavefunctions. We may estimate this amplitude  as ${\cal A}\sim e^{-S_{gravity}}$ where  $S_{gravity}={1\over 16\pi G}\int {\cal R} \sqrt{-g} d^4 x$ and we use  $\sim GM$ for all length scales. This gives
\be
S_{gravity}\sim {1\over G}\int {\cal R}\sqrt{-g} \, d^4 x \sim {1\over G} {1\over (GM)^2}(GM)^4\sim GM^2\sim \Big ( {M\over m_{p}}\Big ) ^2
\ee
Thus ${\cal A}\sim e^{-(M/m_p)^2}$ is indeed tiny. But we must now multiply by the {\it number} of states that we can tunnel to, and this is given by ${\cal N}\sim e^{S_{bek}}$ where
\be
S_{bek}={A\over 4G}\sim {(GM)^2\over G}\sim \Big ( {M\over m_{p}}\Big ) ^2
\ee

\begin{figure}[htbp]
\begin{center}
\includegraphics[scale=.38]{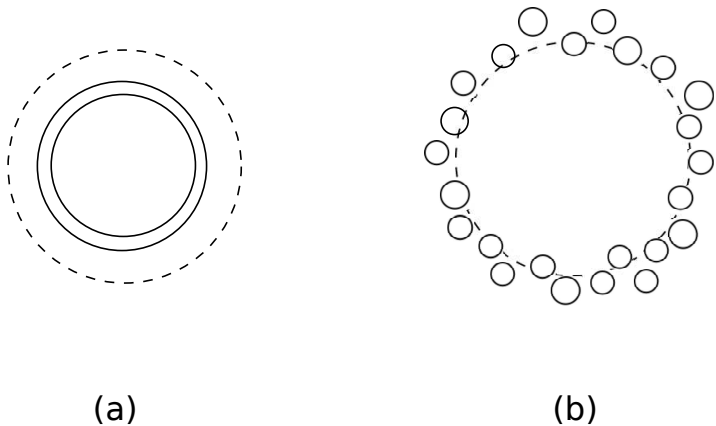}
\caption{{The wavefunction of a collapsing shell tunnels to `fuzzball' states, altering the classically expected dynamics.}}
\label{fthree}
\end{center}
\end{figure}

 Thus the smallness of the tunneling amplitude is offset by the remarkably large degeneracy of states that the black hole has \cite{tunnel}. The wavefunction of the shell tunnels into these `fuzzball' states in a time much shorter than the evaporation time of the hole \cite{rate}, and then these fuzzballs states radiate energy much like any other normal body \cite{cm}. In short, the semiclassical approximation that leads to the standard black hole geometry  gets invalidated by the large measure of phase space over which the wavefunction of the shell can spread.

\begin{figure}[htbp]
\begin{center}
\includegraphics[scale=.48]{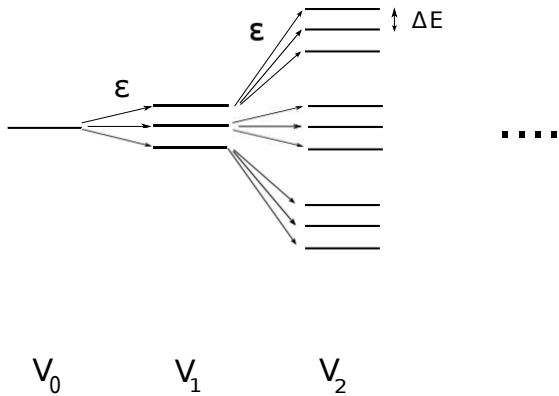}
\caption{{The states at each size $V_k$ can transition with amplitude $\epsilon$ to a band of states in volume $V_{k+1}$, with band spacing $\Delta E$; thus we get a series of `fermi golden rule absorptions' taking us to larger volumes.}}
\label{ffour}
\end{center}
\end{figure}

Now consider the Universe at a time when the volume $V_0$ is very small, but $E$ is very large \cite{cmuniv}. Very few of the  $e^{S_{bek}(E)}$ states allowed at this energy $E$ can fit in our box -- we say that the energy $E$ is `hypercompressed'.  There is a small amplitude $\epsilon$ per unit time for  tunneling between states in $V_0$ and states in a {\it larger} volume $V_1$. Because the entropy of black hole type states is huge,  there are vastly more states in  $V_1$,  and the wavefunction will drift towards the volume $V_1$.  We access even  more states if we expand to a yet larger volume $V_2$, and so on, giving a `push' towards larger volumes. Fig.\ref{ffour} gives a toy quantum mechanical model for this evolution: each state in a given box can transition into a  {\it band} (with energy spacing $\Delta E$) in the next larger box.  The peak of the wavefunction moves through the volumes $V_k$ with \cite{leptonphoton}
\be
k_{peak}(t)\approx {2\pi \epsilon^2\over \Delta E}t
\ee 
Note that  this expansion is in {\it addition} to any expansion rate obtained from the classical gravity equations, since it is generated by the phase space measure describing the degeneracy of states. Note that  the degeneracy of states is always a factor in quantum systems, but what makes it significant for our macroscopic problem is the enormously large entropy of black hole type states.

Obtaining precise rates of expansion from this physics will have to await the extraction  of the parameters $\epsilon, \Delta E$ from string theory. 
But qualitatively, we can see two regimes where this extra `push' can be relevant:

(i) In the early Universe,  matter can be very `hyper-compressed', so we may get  an analogue of inflation.

(ii) The Cosmological horizon $R_H$ today   holds just about the mass needed to make a black hole with radius $R_H$.  Thus matter is on the border of being hyper-compressed, and we may get order unity corrections to the expansion rate. 

\b

To summarize, we have traditionally made little use of the enormous degeneracy of gravitational states implied by the Bekenstein formula. With the explicit realization of these states in string theory, we find that their large measure $D[g]$ competes with the classical  term $e^{-S}$ in path integrals. This correction resolves the information paradox, and  may also provide  Cosmological effects that avoid the need for  fine tuned inflation potentials or anthropic arguments that must tune $\Lambda$ to $10^{-120}m_p^4$.

\section*{Acknowledgements}

This work was supported in part by DOE grant DE-FG02-91ER-40690.

\b

\b

\end{document}